\documentclass{emulateapj}
\usepackage{graphicx}
\usepackage{color}
\usepackage{amsmath}

\newcommand{\Fewbody}{{\em Fewbody\/}}

\newcommand{\Starlab}{{\em Starlab\/}}

\begin{document}

\title{Dynamical Interactions of Planetary Systems in Dense Stellar Environments}
\shorttitle{Dynamical Interactions of Planetary Systems}
\submitted{Accepted for publication in \apj}
\author{John M. Fregeau\altaffilmark{1}, Sourav Chatterjee\altaffilmark{2}, \& Frederic A. Rasio\altaffilmark{3}}
\shortauthors{FREGEAU, ET AL.}
\affil{Department of Physics and Astronomy, Northwestern University, Evanston, IL 60208}
\altaffiltext{1}{fregeau@alum.mit.edu}
\altaffiltext{2}{s-chatterjee@northwestern.edu}
\altaffiltext{3}{rasio@northwestern.edu}

\begin{abstract}
We study dynamical interactions of star--planet binaries with other single stars.
We derive analytical cross sections for all possible outcomes, and confirm them
with numerical scattering experiments.  We find that a wide mass ratio
in the binary introduces a region in parameter space that is inaccessible to
comparable-mass systems, in which the nature of the dynamical interaction is 
fundamentally different from what has traditionally been considered in the 
literature on binary scattering.  We study the properties of the planetary
systems that result from the scattering interactions for all regions of parameter
space, paying particular attention to the location of the ``hard--soft'' boundary.
The structure of the parameter space turns out to be significantly richer
than a simple statement of the location of the ``hard--soft'' boundary would imply.
We consider the implications of our findings, calculating characteristic lifetimes
for planetary systems in dense stellar environments, and applying the results
to previous analytical studies, as well as past and future observations.
Recognizing that the system PSR B1620-26 in the globular cluster M4 lies in the
``new'' region of parameter space, we perform a detailed analysis quantifying
the likelihood of different scenarios in forming the system we see today.
\end{abstract}

\keywords{stellar dynamics --- planetary systems --- globular clusters: general --- 
  methods: $n$-body simulations --- pulsars: individual (1620-26)}

\section{Introduction}\label{sec:intro}

The subject of planets in dense stellar systems such as globular clusters
came to life with the discovery that the binary pulsar PSR B1620-26 in the 
globular cluster M4 is orbited by a distant companion of approximately Jupiter
mass \citep[][and references therein]{1993ASPC...36...11B,1999ApJ...523..763T,2003Sci...301..193S}.  The presence
of such a system near the core of M4 is, at first, puzzling for several reasons.
For example, the metallicity of M4 is low ($0.05 Z_\sun$), making the formation of such a planet
via the standard core accretion model very unlikely, and sparking the creation of
alternative formation theories \citep[][and references therein]{2004MNRAS.355.1244B}.
In addition, the planet is on such a wide orbit about its host that the system should have been destroyed
by dynamical interactions on a very short timescale ($\sim 10^8\,{\rm yr}$) in the
moderately high density environment of M4's core, implying that it must have been
formed recently via an alternative mechanism such as a dynamical interaction involving a binary.

Campaigns searching for ``hot Jupiter'' planets (with
orbital periods $\la 10\,{\rm d}$) in 47 Tuc have yielded null detections, instead
placing upper limits on the hot Jupiter frequency there
\citep{2000ApJ...545L..47G,2005ApJ...620.1043W}.  It now appears that the lack
of hot Jupiters detected in 47 Tuc is probably more closely connected with the low
metallicity of the environment, and not simply due to destruction via
dynamical encounters in the high density stellar environment.
This result is consistent with the strong correlation between
host star metallicity and planet frequency in observed systems \citep{2005ApJ...622.1102F}.
Searches for planets via transits in the higher metallicity environments
of open clusters are currently underway\footnote{see http://star-www.st-and.ac.uk/\~{}kdh1/transits/table.html}.

A key ingredient in the formation and evolution of planetary systems in dense stellar 
environments that is absent in lower density environments is dynamics.  If either
the stellar density is large enough or the planetary system is wide enough, it is likely
to undergo a strong dynamical scattering interaction with another star or binary, perturbing
its properties or possibly destroying it in the process.  
The importance of dynamics for wide mass ratio binaries cannot be overstated, 
especially when one considers the 
wealth of low-mass objects such as brown dwarfs that are known to exist in open
and globular clusters, and may be in binaries.  In addition, the importance of dynamics 
as early as the stage of star formation is now becoming clear.  There is now convincing
evidence that a distant encounter with another star is responsible for the currently
observed properties of distant Solar System objects such as Sedna
\citep{2004Natur.432..598K,2004AJ....128.2564M,2005Icar..177..246K}.
The object HD 188753 is a hierarchical triple star system composed of a main-sequence
star in an $\approx 12.3\,{\rm AU}$ orbit about a compact main-sequence binary.  What makes
this system interesting is that the single star hosts a hot Jupiter in orbit about it,
with a semimajor axis of $\approx 0.04\,{\rm AU}$ \citep{2005Natur.436..230K}.  
The current configuration of the system
is puzzling, since the binary companion would truncate the initial disk out of which
the planet could have formed at a radius much smaller than that required by either 
the core accretion or gravitational fragmentation paradigms.  It is now clear
that the current configuration of HD 188753 must have resulted from a dynamical scattering interaction 
in a moderately dense stellar environment \citep{pfahl2005,spz2005}.

The types of dynamical interactions we have been discussing
have been studied in great detail and volume since at least the mid-1970s
for the case of comparable masses 
\citep[][and references therein]{1975AJ.....80..809H,1975MNRAS.173..729H,2003gmbp.book.....H}.
The case of disparate masses in the binary (such as for a planetary system)
has been treated in less detail 
\citep{1984AJ.....89.1559H,1989AJ.....98.1069H,1992ApJ...399L..95S},
but with considerable attention paid to the survivability of planetary systems
in particular \citep{1998ApJ...508L.171L,2001MNRAS.322..859B,2001MNRAS.324..612D}.
To the best of our knowledge, there has not yet been a comprehensive treatment of dynamical
scattering interactions with wide mass ratio systems in the literature, as previous work has
typically been concerned with a numerical treatment of a certain aspect of the problem.  This paper is
an attempt to remedy this situation by carefully treating a wide range of parameter
space both analytically and numerically, and identifying physically
distinct regions of the space, with an emphasis on application to
physically interesting problems.

In particular, the questions we would like to answer in this paper
are the following: for a dynamical interaction
between a star--planet system and another star, under what conditions does the interaction
result in: 1) preservation of the planetary system, in the sense that the planet remains
in orbit about its parent star, even if the orbit is perturbed; 2) exchange of the incoming
star for the planetary system's host star; 3) destruction of the planetary
system, in the sense that the planet is not bound to either star after the interaction; and 4) 
physical collisions involving two or more of the three objects?  
For the outcome in question 2,
we would also like to know the resulting properties of the planetary system's orbit.  
In particular, we would like to know the regimes in which the planetary system orbit 
shrinks and those in which it widens.

\begin{deluxetable*}{ccc}
  \tablecaption{Possible outcomes of a dynamical scattering interaction between a planetary
    system, `[$M_1$~$m$]', and a single star, `$M_2$'.\label{tab:outcomes}}
  \tablehead{
    \colhead{outcome} & \colhead{description} & \colhead{nickname}
  }
  \startdata
  \null[$M_1$~$m$]~$M_2$ &preservation &``pres''\\
  \null[$M_2$~$m$]~$M_1$ &exchange host star for incoming star &``ex$\_$s''\\
  \null[$M_1$~$M_2$]~$m$ &exchange planet for incoming star &``ex$\_$p''\\
  \null$M_1$~$M_2$~$m$ &ionization &``ion''\\
  \hline
  \null[$M_i$:$m$~$M_j$] &collision of planet with one star, resulting in binary &\\
  \null$M_i$:$m$~$M_j$ &collision of planet with one star, resulting in ionization &\\
  \null[$M_i$:$M_j$~$m$] &collision of two stars, resulting in binary &\\
  \null$M_i$:$M_j$~$m$ &collision of two stars, resulting in ionization &\\
  \hline
  \null$M_i$:$M_j$:$m$ &collision of all objects &
  \enddata
  \tablecomments{The symbol `:' denotes a physical collision.}
\end{deluxetable*}

In the language of binary scattering, the interaction of a binary planetary system with a 
single star can be written symbolically as `[$M_1$~$m$]~$M_2$', where the $M_i$ represent
the stars, and $m$ represents the planet.  The possible outcomes can be enumerated succinctly
as shown in Table~\ref{tab:outcomes}.  The outcome ``pres'' is preservation
of the planetary system, and can result either from a strong interaction or a weak fly-by.
The planetary system can have shrunk or widened in the process.  The outcome ``ex$\_$s''
is an exchange in which the host star of the planetary system
is exchanged for the incoming star.  The outcome ``ex$\_$p'' is an exchange
in which the two stars remain bound, ejecting the planet.  The outcome ``ion''
leaves all objects (stars and planet) unbound from each other.  The remaining outcomes
listed in the table are various permutations involving one or more collisions.
The outcome of a stable hierarchical triple is omitted from the table since it is classically
forbidden \citep{2003gmbp.book.....H}.  It is possible to be formed as a result of a binary--single
interaction if tidal effects are present.  However, we ignore such effects in this paper.

Clearly, case 1 above corresponds to outcome ``pres'', case 2 corresponds
to outcome ``ex$\_$s'', and case 3 corresponds to outcomes ``ex$\_$p'' and ``ion''.  
The outcomes corresponding to case 4 are the remaining listed in the table.  In this paper
we address the questions posed above by calculating cross sections for the processes
listed, as well as characteristic lifetimes.  In section \ref{sec:analytic} we treat as much
of the problem as we can analytically.  In section \ref{sec:numeric} we confirm
our analytical predictions with numerical experiments.  
In section \ref{sec:hardsoft} we determine under what conditions the planetary
system will shrink or expand as the result of an exchange encounter, and the average
amount by which it shrinks or expands.  In section \ref{sec:implications}
we discuss the implications of the results in the context of current literature
and observations.  In section \ref{sec:1620} we quantify the likelihood of the possible
dynamical formation scenarios of the planetary system PSR B1620-26 in the globular cluster M4.  
Finally, in section \ref{sec:summary} we summarize and conclude.

\section{Analytical Considerations}\label{sec:analytic}

\begin{figure}
  \begin{center}
    \includegraphics[width=\columnwidth]{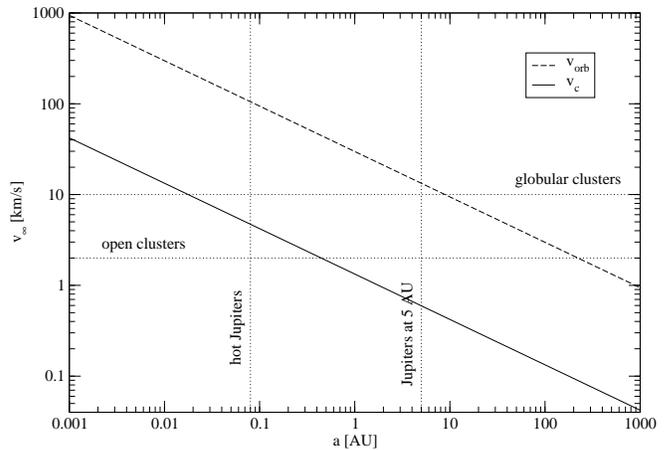}
    \caption{Plot of the critical velocity $v_c$ and orbital speed $v_{\rm orb}$ for a planetary system
      composed of a $1 M_\sun$ star and a $10^{-3} M_\sun$ planet in an orbit with semimajor axis
      $a$, encountering a $1 M_\sun$ star with relative velocity at infinity $v_\infty$.  Dotted lines
      represent characteristic velocity dispersions for open and globular clusters, and orbital
      sizes for hot Jupiters and Jupiters at $5\,{\rm AU}$, as shown.\label{fig:phasespace}}
  \end{center}
\end{figure}

As described above, we symbolically label the interaction between a planetary system
and a single star as `[$M_1$~$m$]~$M_2$'.  We also use the symbols $M_i$ and $m$
for the masses of the bodies, along with $a$ for the planetary system semimajor axis, 
$e$ for its eccentricity, $v_\infty$
for the relative velocity at infinity between the planetary system and single star,
and $b$ for the impact parameter.  If one treats all bodies as point particles, and only 
considers their masses for the time being, there are only two velocity scales in the
problem.  The first is what has historically been called the critical velocity, which 
is the value of $v_\infty$ for which the total
energy of the binary--single system is zero, and is given by
\begin{equation}\label{eq:vcrit}
  v_c = \left(\frac{GM_1m}{\mu a}\right)^{1/2} \, ,
\end{equation}
where $\mu=(M_1+m)M_2/(M_1+M_2+m)$ is the reduced mass of the binary--single star system.
The second is the orbital speed of the planet relative to its parent star, and is given by
\begin{equation}\label{eq:vorb}
  v_{\rm orb} = \sqrt{\frac{G(M_1+m)}{a}} \, ,
\end{equation}
for a circular orbit.  
Looking at eqs.~(\ref{eq:vcrit}) and (\ref{eq:vorb}), we see that if all the masses
in the problem are comparable, then $v_c \approx v_{\rm orb}$.  If, on the other
hand, one of the members of the binary system is much less massive than the other member (as is
the case for a planetary system),
and we take $M_1=M_2\equiv M$, 
and $q\equiv m/M \ll 1$, then we can approximately write 
$v_c\approx\sqrt{q}v_{\rm orb}$.  For a typical planetary system, $q \sim 10^{-3}$,
and the two velocities differ by more than an order of magnitude.  Fig.~\ref{fig:phasespace}
plots the two velocities for a planetary system with $q=10^{-3}$ and $M=M_\sun$ as a function
of $a$.  In essence, introducing a mass ratio in the binary opens up a strip in parameter
space (between the two lines in Fig.~\ref{fig:phasespace}) in which the physical 
character of dynamical scattering interactions is fundamentally different 
from that of the comparable-mass case.

In the literature on binary scattering involving stars of comparable
mass, $v_c$ is the velocity delimiting the border between ``hard'' and ``soft'' 
binaries \citep{1975MNRAS.173..729H}.  If $v_\infty < v_c$, the binary is in the hard regime, and will on average
harden ($a$ will get smaller) as a result of an encounter.  On the 
other hand, if $v_\infty > v_c$, the binary is in the soft regime, and will on average
soften ($a$ will get larger) as a result of an encounter---that is, if the binary is
not destroyed completely.  However, \citet{1989AJ.....98.1069H} and \citet{1990AJ.....99..979H} 
suggest that when considering binaries with disparate masses, the boundary is more 
accurately given by $v_\infty=v_{\rm orb}$, and not by $v_\infty=v_c$, and suggest
the use of the terminology ``fast--slow boundary'' rather than ``hard--soft boundary'', 
since it is the relative speed of the binary members that is physically relevant.  In 
the literature considering the survivability of planetary systems (and high mass ratio systems
in general) in dense stellar environments, 
both $v_c$ and $v_{\rm orb}$ have been used as the single characteristic velocity delimiting the
boundary between ionization on the high-velocity side and hardening
on the low-velocity side.
However, as we will see below, for planetary systems the hard--soft boundary lies at $v_\infty=v_c$, 
while the characteristic timescale for the survivability of a planetary system in a dense environment
drops markedly only at $v_\infty=v_{\rm orb}$.

Having written down the two characteristic velocities in the problem, thus dividing parameter space into three
distinct regions (as shown in Figure~\ref{fig:phasespace}), the question naturally arises: What is the 
physical character of the scattering interactions in each of the three regions?  For $v_\infty<v_c$, the total
energy of the binary--single system is negative.  We thus expect that for any ``strong'' encounter (defined
approximately as one in which the classical pericenter distance, $r_p$, between the single star and the planetary system
is of order $a$), the interaction will be resonant, in the sense that it will
survive for many orbital times \citep{2003gmbp.book.....H}.  It will behave as a small star cluster,
most likely ejecting its lightest member---in this case the planet---yielding the outcome
``ex$\_$p (res)''.  For $r_p \ga a$, the likely outcome is a ``direct'' exchange of the incoming 
star for the planet (``ex$\_$p'').  We thus expect 
$\sigma_{\rm ex\_p}=\sigma_{\rm ex\_p (res)}+\sigma_{\rm ex\_p (non-res)}$ to 
be the dominant cross section for $v_\infty < v_c$.
For $v_\infty>v_{\rm orb}$, the interaction is impulsive, since the timescale of the interaction is a small
fraction of the binary orbital period.  We thus expect that any ``strong'' interaction will ionize
the system, making $\sigma_{\rm ion}$ the dominant cross section in this regime.
For $v_c < v_\infty < v_{\rm orb}$ the timescale of the interaction is comparable to the orbital period,
so the outcome is not a priori obvious, and must be treated numerically.

Below, as far as it is possible, we calculate using analytical techniques the cross sections for each
outcome in each of the three regions of parameter space.

\subsection{$v_\infty < v_c$}

First, since for $v_\infty < v_c$ the total energy of the binary--single system is negative, it is
clear that ionization is classically forbidden.  Thus
\begin{equation}
  \sigma_{\rm ion} = 0 \, .
\end{equation}
We should also mention that for the outcome of preservation the only method we have for determining
if an interaction was ``strong'' is whether or not it was resonant.  Thus it should be clear
that for every region of parameter space in Figure~\ref{fig:phasespace}
\begin{equation}
  \sigma_{\rm pres (non-res)} \rightarrow\infty \, ,
\end{equation}
since every very distant passage of the single star by the binary preserves it via a non-resonant
interaction.

\citet{1996ApJ...467..359H} considered interactions of ``hard'' binaries (in the sense
that $v_\infty < v_c$) with single stars for a wide range of possible mass ratios in the binary 
and between the binary members and the incoming single star.  Using analytical techniques,
they calculated the scaling of the cross sections for each possible outcome.  
They then fit to numerically calculated cross sections to determine the weakly mass-dependent
``coefficients'' on each cross section.
From their eqs.~(13) and (14) with $m_2 \approx m_3$ (since $M_1 \approx M_2$ here), we see that 
\begin{equation}\label{eq:z}
  \sigma_{\rm ex\_p (res)} \approx \sigma_{\rm ex\_p (non-res)} \, .
\end{equation}
The cross section for non-resonant exchange of the planet for the incoming star
is given by their eq.~(17) (divided by 2 since $\sigma_{\rm ex\_p (res)} \approx \sigma_{\rm ex\_p (non-res)}$):
\begin{multline}\label{eq:a}
  \sigma_{\rm ex\_p (non-res)} \approx
  \frac{\pi G M_t a}{4 v_\infty^2} \left(\frac{M_1+M_2}{M_t}\right)^{1/6}\\
  \times\left(\frac{M_2}{m+M_2}\right)^{7/2} \left(\frac{M_1+m}{M_t}\right)^{-1/3} \left(\frac{M_2+m}{M_t}\right)\\
  \times\exp\left(\sum_{m,n}a_{mn}\left(\frac{m}{M_1+m}\right)^m \left(\frac{M_2}{M_t}\right)^n\right)\, ,
\end{multline}
where $M_t=M_1+m+M_2$ is the total mass of the binary--single system, the exponential term
is the ``coefficient'' fit to the numerical results, and the $a_{mn}$ are given in their Table 3.

\begin{figure}
  \begin{center}
    \includegraphics[width=\columnwidth]{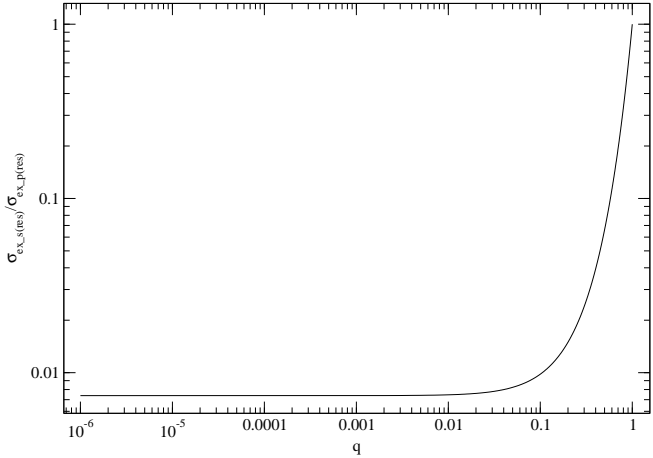}
    \caption{Plot of the ratio of the cross section for resonant exchange in which the planet's original host star
      is ejected, to that in which the planet is ejected, $\sigma_{\rm ex\_s (res)}/\sigma_{\rm ex\_p (res)}$,
      as a function of the binary mass ratio $q\equiv m/M$.  Here we have assumed that both stars have mass $M$.  Note how,
      as $q$ is decreased from unity, the test-particle limit is reached relatively quickly, at $q\approx 0.1$.
    \label{fig:sigma_res_ratio}}
  \end{center}
\end{figure}

The cross section for direct exchange of the incoming star for the host star
is also given by their eq.~(17), with the appropriate permutation of masses:
\begin{multline}\label{eq:b}
  \sigma_{\rm ex\_s (non-res)} \approx
  \frac{\pi G M_t a}{2 v_\infty^2} \left(\frac{M_2+m}{M_t}\right)^{1/6}\\
  \times\left(\frac{M_2}{M_1+M_2}\right)^{7/2} \left(\frac{M_1+m}{M_t}\right)^{-1/3} \left(\frac{M_1+M_2}{M_t}\right)\\
  \times\exp\left(\sum_{m,n}a_{mn}\left(\frac{M_1}{M_1+m}\right)^m \left(\frac{M_2}{M_t}\right)^n\right)\, .
\end{multline}
The resonant cross section for this process, $\sigma_{\rm ex\_s (res)}$,
can be calculated approximately by statistical techniques.  Since for a resonant interaction we expect the three-body
system to behave like a small star cluster, we can assume that the less massive body (the planet)
and the more massive bodies (the stars) reach energy equipartition, yielding the approximate equality
\begin{equation}
  m v_{\sigma,m}^2 = M v_{\sigma,M}^2 \, ,
\end{equation}
where $v_{\sigma,m}$ is the one-dimensional velocity ``dispersion'' of the planet, $v_{\sigma,M}$ is the one-dimensional
velocity dispersion of the stars, and we let $M_1=M_2=M$.  Assuming that the three body system reaches 
thermal equilibrium on a timescale shorter than the interaction timescale, 
each species obeys a Maxwellian distribution:
\begin{equation}
  f(v, n, v_\sigma) = 4\pi v^2\frac{n}{(2\pi v_\sigma^2)^{3/2}} e^{-v^2/2v_\sigma^2} \, ,
\end{equation}
where $n$ is the number density of the species, $v_\sigma$ its one-dimensional velocity dispersion, and 
$v$ its speed.  To find the ratio of the probability for ejecting one of the stars to ejecting 
the planet, one simply integrates the velocity distribution for each species from 
the escape velocity, $v_{\rm esc}$, to infinity and considers the ratio.  Using the fact that
$v_{\rm esc} \approx 2 \sqrt{3} v_\sigma$, where $v_\sigma \approx v_{\sigma,M}$ is the one-dimensional velocity
dispersion of the full three-object system, we have
\begin{equation}\label{eq:c}\begin{split}
  \frac{\sigma_{\rm ex\_s (res)}}{\sigma_{\rm ex\_p (res)}} &= 
  \frac{\int_{2\sqrt{3}v_\sigma}^\infty f(v, n, v_\sigma) dv}{\int_{2\sqrt{3}v_\sigma}^\infty f(v, n, v_\sigma/\sqrt{q}) dv}\\
  &= \frac{2e^{-6}\sqrt{\frac{6}{\pi}}+{\rm erfc}(\sqrt{6})}{2e^{-6q}\sqrt{\frac{6q}{\pi}}+{\rm erfc}(\sqrt{6q})} \, ,
\end{split}\end{equation}
where ${\rm erfc}(z) = 1-{\rm erf}(z)$ is the complement of the error function.  
Note that there is no factor of two in the numerator since the two stars in the system are distinguishable,
in the sense that we distinguish between ``ex$\_$s (res)'' and ``pres (res)''.
This function
is plotted in Fig.~\ref{fig:sigma_res_ratio}.  Note how,
as $q$ is decreased from unity, the test particle limit is reached relatively quickly, at $q\approx 0.1$.  The
same qualitative result is found in \citet{2004MNRAS.352....1F}, as shown in their Figure~10.

Since in a resonant interaction the system loses all memory of its initial configuration, we expect
that for $M_1=M_2=M$, the probability of preserving the system (and ejecting the incoming star) in a resonant interaction
is equal to that of ejecting the host star and keeping the incoming star:
\begin{equation}\label{eq:d}
  \sigma_{\rm pres (res)} = \sigma_{\rm ex\_s (res)} \, .
\end{equation}

Finally, we remark on the likelihood of collisions in this regime.  \citet{2004MNRAS.352....1F}
have studied in detail the probability of collisions during binary interactions, including
binaries with very small mass ratios, and found that for systems in which the radius of the
low-mass member is independent of its mass (as is approximately the case for planets), the 
test particle limit is reached at $q\approx 0.1$ (see Figure~10 of that paper).  
In other words, the collision cross section is independent of the mass of the low-mass member 
below $q\approx 0.1$.  Furthermore, the collision cross section is roughly as large as the 
strong interaction cross section.  Therefore, we expect the combined cross section for any outcome
involving a collision to be on the same order as $\sigma_{\rm ex\_p (res)}$.  This estimate
says nothing about the relative frequency of the different outcomes involving collisions, 
which requires numerical calculations to infer, and which we do not treat in this paper.
However, we do consider collisions briefly in the discussion of PSR B1620-26 in section \ref{sec:1620}.

\subsection{$v_\infty > v_{\rm orb}$}

In the regime $v_\infty > v_{\rm orb}$, since $v_\infty > v_c$, the total energy of the system
is positive, so resonant scattering is forbidden:
\begin{equation}
  \sigma_{\rm pres (res)} = \sigma_{\rm ex\_s (res)} = \sigma_{\rm ex\_p (res)} = 0 \, .
\end{equation}
Since $v_\infty > v_{\rm orb}$, the timescale of the scattering interaction is typically
much smaller than the orbital period (unless $r_p \gg a$).  In this case the interaction
is in the impulsive limit, and can be handled quite efficiently with analytical techniques
\citep{1983ApJ...268..342H}.  From eq.~(5.1$^\prime$) of \citet{1983ApJ...268..342H}, the
ionization cross section is 
\begin{equation}\label{eq:e}
  \sigma_{\rm ion} = \frac{40}{3}\pi a G \frac{M_2^2}{M_1+m} \frac{1}{v_\infty^2} \, .
\end{equation}

Exchange in the impulsive regime occurs when the incoming star, $M_2$, passes much closer
to one member of the binary than the other, and undergoes a near-180$^\circ$ encounter in
which all of its momentum is transferred to that member.  In order
for $M_2$ to remain bound in the binary, it must receive nearly zero recoil in the frame
of the remaining binary member, requiring that the mass of $M_2$ be nearly equal to
the mass of the binary member it displaces \citep{1983ApJ...268..342H}.  Thus exchange
of the planet for the incoming star, $\sigma_{\rm ex\_p (non-res)}$,
is vanishingly small unless $M_2 \approx m$.  Exchange of the planet's host star for the 
incoming star, on the other hand, is more likely since the stars' masses
are more likely to be comparable.  From eq.~(5.2$^\prime$) of \citet{1983ApJ...268..342H}, 
the cross section for exchange of the planet's host star for the incoming star is given by
\begin{equation}\label{eq:f}
  \sigma_{\rm ex\_s (non-res)} = \frac{20}{3}\pi \frac{1}{a} G^3 M^2 (M+m) \frac{1}{v_\infty^6} \, ,
\end{equation}
with $M_1=M_2=M$.

\subsection{$v_c < v_\infty < v_{\rm orb}$}

In the regime $v_c < v_\infty < v_{\rm orb}$, the interaction is neither resonant nor
impulsive, so the techniques used above cannot be used here.  For $q \ll 1$ the interaction
is essentially the hyperbolic restricted three-body problem, but since the timescale of the
interaction is of order a few orbital periods, the standard techniques of perturbation theory
are of limited use.  Instead, one must appeal to numerical methods to discern the nature
of the interactions.  It is possible, however, to make order of magnitude estimates of 
the relevant cross sections.  Whenever $M_2$ passes within $r_p \la a$ 
of the binary, its force on the planet is comparable to that of $M_1$, in which
case the orbit of the planet should be sufficiently perturbed so as to unbind it from the host
star and either eject it from the system or leave it bound to $M_2$.  Thus we estimate that
the total cross section $\sigma_{\rm ion}+\sigma_{\rm ex\_s (non-res)}$ should be roughly
equal to the strong interaction cross section.  In the absence of any more detailed information,
we simply assume $\sigma_{\rm ion}\approx\sigma_{\rm ex\_s (non-res)}$, yielding
\begin{equation}\label{eq:g}
   \sigma_{\rm ion} \sim \sigma_{\rm ex\_s (non-res)} \sim \pi a \frac{G M_t}{v_\infty^2} \, ,
\end{equation}
since gravitational focusing dominates.  Since in this regime the planet is effectively a 
test particle, $\sigma_{\rm ex\_p (non-res)}$ is vanishingly small.

\section{Numerical Confirmation}\label{sec:numeric}

To confirm the analytical predictions of the previous section, we have
calculated the cross sections by numerically integrating many scattering interactions.  We
have used the \Fewbody\ numerical toolkit to perform the scattering interactions
\citep{2004MNRAS.352....1F}.  For calculating the cross sections, we use a generalization
of the technique described in \citet{1996ApJ...467..348M} in which for each annulus
in impact parameter we perform only one scattering interaction, instead of the many
required by their technique.  This allows us to more precisely calculate the uncertainties
in the cross sections, in particular yielding accurate asymmetric error bars.  To handle
the small number statistics, we use the tables in \citet{1986ApJ...303..336G}.

\begin{figure}
  \begin{center}
    \includegraphics[width=\columnwidth]{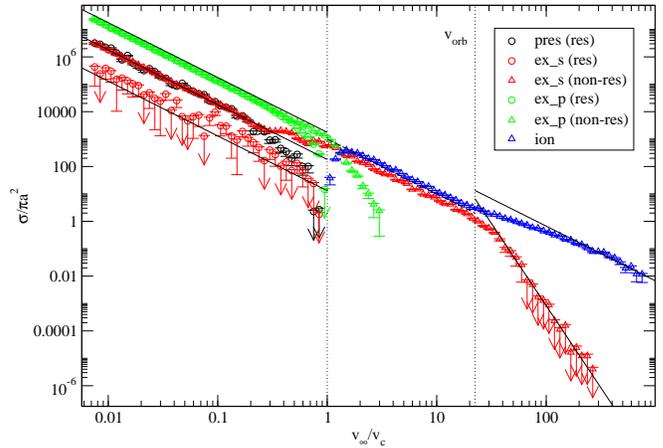}
    \caption{The dimensionless cross sections, $\sigma/\pi a^2$, for the outcomes involving no collisions
      in Table~\ref{tab:outcomes}, as a function of $v_\infty/v_c$ for scattering interactions between a
      planetary system with mass ratio $q=m/M=10^{-3}$ and eccentricity $e=0$, and a star of mass $M$.  
      Points plotted are as described in the legend.  Circles represent
      resonant interactions, while triangles represent non-resonances (``direct'' 
      interactions).  The error bars plotted are one-sigma.
      The vertical dotted lines are the critical velocity and orbital speed.
      The solid lines are the analytically predicted cross sections from section \ref{sec:analytic}, 
      while the dashed line is the order of magnitude estimated cross section from the same section.
      \label{fig:sigmas}}
  \end{center}
\end{figure}

Figure~\ref{fig:sigmas} shows the dimensionless cross sections, $\sigma/\pi a^2$, for 
the outcomes involving no collisions
in Table~\ref{tab:outcomes}, as a function of $v_\infty/v_c$ for a planetary system
with mass ratio $q=10^{-3}$ and $e=0$, and with $M_1=M_2=M$.  
Points plotted are as described in the legend.  Circles represent
resonant interactions, while triangles represent non-resonances (``direct'' 
interactions).  The error bars plotted are one-sigma.  
The vertical dotted lines are the critical velocity and orbital speed.
The solid lines are the analytically predicted cross sections 
from section \ref{sec:analytic}, while the dashed line is the order of magnitude estimated cross 
section from the same section.  We have calculated the same cross sections using 
\Starlab\ \citep{1996ApJ...467..348M,2001MNRAS.321..199P}, and the agreement is excellent.
Note that with the exception of section~\ref{sec:1620}, all calculations presented in this
paper are performed with $e=0$.  Previous results show that most cross sections of interest
do not depend strongly on eccentricity \citep{1983ApJ...268..319H}.  Indeed, we have calculated
the cross sections in Figure~\ref{fig:sigmas} using a thermal eccentricity distribution and
find only a slight increase in the resonant cross sections (and corresponding decrease in the 
non-resonant cross sections) relative to the $e=0$ case.

Looking first at the region $v_\infty > v_{\rm orb}$, we see $\sigma_{\rm ion}$ plotted
in blue and $\sigma_{\rm ex\_s (non-res)}$ plotted in red, along with the corresponding
analytically predicted cross sections (eqs.~[\ref{eq:e}] and [\ref{eq:f}], respectively) 
plotted as solid lines.  The agreement is clearly quite excellent.

Next, looking at the region $v_\infty < v_c$, we see $\sigma_{\rm ex\_p (res)}$ (circles)
and $\sigma_{\rm ex\_p (non-res)}$ (triangles) plotted in green, with the corresponding
analytically predicted result of eq.~(\ref{eq:a}) (the two cross sections are expected to be roughly
equal, from eq.~[\ref{eq:z}]) plotted as a solid line.  $\sigma_{\rm ex\_p (res)}$
and $\sigma_{\rm ex\_p (non-res)}$ appear to be roughly equal, with $\sigma_{\rm ex\_p (res)}$
larger by $\sim 10\%$ for $v_\infty \ll v_c$, while the analytical prediction
is roughly $\sim 50\%$ larger than the numerical result.  Plotted in red triangles
is $\sigma_{\rm ex\_s (non-res)}$, along with the corresponding analytical result
from eq.~(\ref{eq:b}) plotted as a solid line.  Clearly the agreement is superb.  Plotted in red circles
is the resonant cross section $\sigma_{\rm ex\_s (res)}$, with the analytical
result from eq.~(\ref{eq:c}) plotted as a solid line.  Although the statistics
on this very small cross section are not as good as the other cross sections for
$v_\infty < v_c$, it still appears to agree very well with the analytical
prediction.  Finally, plotted in black circles is $\sigma_{\rm pres (res)}$.
Although from eq.~(\ref{eq:d}) above we expect it to be equal to
$\sigma_{\rm ex\_s (res)}$, it is larger by nearly an order of magnitude.
The discrepancy is due to the operational definition of resonance we have used---namely
that the mean-squared separation of all bodies makes more than one minimum, with a
successive minimum counted only if the intervening maximum between it and the previous
minimum is at least twice the value of both minima \citep{1996ApJ...467..348M}.  Clearly
this definition is too liberal, and classifies some preservation outcomes as resonant
when they should not be classified as such.  In fact, requiring that the mean-squared separation
make more than two minima for a resonance decreases $\sigma_{\rm pres (res)}$ by about
an order of magnitude, bringing it into agreement with the analytical prediction and making it
nearly equal to $\sigma_{\rm ex\_s (res)}$.  Requiring that intervening maxima be four times 
(instead of twice) the value of the minima on either side for a successive minimum to be counted has 
no discernible effect on the results.

In the region $v_c < v_\infty < v_{\rm orb}$ resonance is, of course, forbidden.
The cross section $\sigma_{\rm ex\_p (non-res)}$ drops off quickly above $v_\infty = v_c$,
obeying roughly
\begin{equation}\label{eq:h}
  \frac{\sigma_{\rm ex\_p (non-res)}}{\pi a^2} \approx 1.5\times 10^3 \left(\frac{v_\infty}{v_c}\right)^{-6} \, ,
\end{equation}
for the case $M_1=M_2=M$, and $q=m/M=10^{-3}$.  The cross sections for ionization and exchange of the 
incoming star for the host star (plotted in blue and red, respectively) are of the same order
of magnitude, and only slightly different from the order of magnitude estimate in eq.~(\ref{eq:g})
(shown in the dashed line).  In particular, for the bulk of the region,
$\sigma_{\rm ion} \approx 2 \sigma_{\rm ex\_s (non-res)}$, and $\sigma_{\rm ion}$ is
$\sim 20\%$ larger than the order of magnitude estimate, yielding
\begin{equation}\label{eq:i}
  \sigma_{\rm ion} \approx 1.2 \pi a \frac{G M_t}{v_\infty^2} \, ,
\end{equation}
and
\begin{equation}\label{eq:j}
  \sigma_{\rm ex\_s (non-res)} \approx 0.6 \pi a \frac{G M_t}{v_\infty^2} \, ,
\end{equation}
again, for the case $M_1=M_2=M$, and $q=m/M=10^{-3}$.  However, as discussed
above in section \ref{sec:analytic}, we expect the test-particle limit to be
reached at $q \approx 10^{-1}$, so the results here should be valid for 
all $q \la 10^{-1}$.

\section{Properties of Exchange Systems and the Hard--Soft Boundary}\label{sec:hardsoft}

\begin{figure}
  \begin{center}
    \includegraphics[width=\columnwidth]{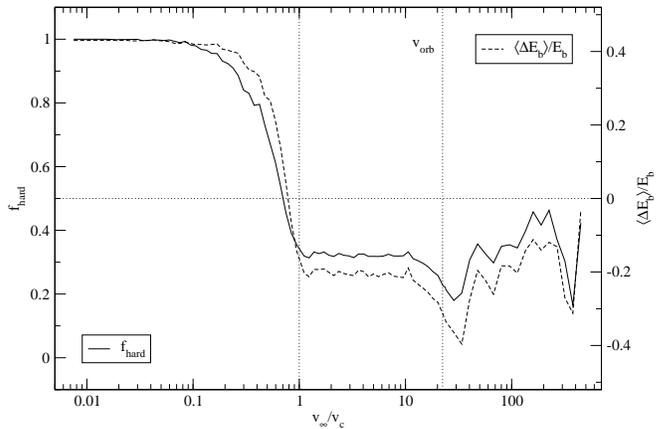}
    \caption{Statistics on the final planetary systems formed via 
      non-resonant exchange in which the planet's
      original host star is ejected (``ex$\_$s (non-res)''), as a function of $v_\infty/v_c$, for the case
      $M_1=M_2=M$ and $q=10^{-3}$.  The fraction of final planetary systems that have 
      hardened (semimajor axis decreased), $f_{\rm hard}$, is shown in the solid line.  
      The average change in the binding energy of the planetary system, 
      $\langle \Delta E_b\rangle/E_b=\langle GMm/2a^\prime- GMm/2a\rangle/(GMm/2a)=a\langle 1/a^\prime\rangle-1$, 
      is shown in the dashed line.  The vertical dotted lines are the critical velocity and the orbital 
      speed.  The horizontal dotted line represents the value 0.5 for $f_{\rm hard}$ and 0 
      for $\langle \Delta E_b \rangle/E_b$.  The hard--soft boundary lies where $f_{\rm hard}$ crosses 
      $0.5$, and $\langle \Delta E_b \rangle/E_b$ crosses $0$, which clearly and robustly occurs at 
      $v_\infty \approx 0.75 v_c$.\label{fig:hardsoft}}
  \end{center}
\end{figure}

\begin{figure}
  \begin{center}
    \includegraphics[width=\columnwidth]{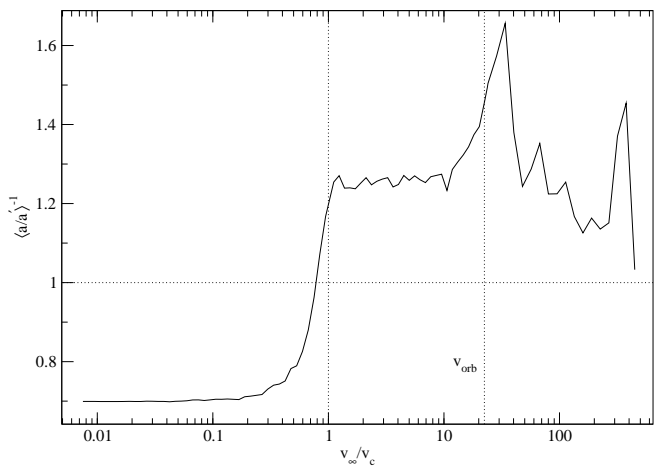}
    \caption{The quantity $\langle a/a^\prime\rangle^{-1}$, the inversely-weighted 
      average of the ratio of the planetary system's
      final semimajor axis $a^\prime$ to its initial value $a$,
      as a function of $v_\infty/v_c$ for the planetary systems formed via 
      non-resonant exchange in which the planet's
      original host star is ejected (``ex$\_$s (non-res)''),
      for the case $M_1=M_2=M$ and $q=10^{-3}$.  The vertical dotted lines show $v_c$
      and $v_{\rm orb}$, while the horizontal dotted line shows the value 1.\label{fig:dascale}}
  \end{center}
\end{figure}

Having determined the character of the scattering interactions in each region of 
parameter space shown in Fig.~\ref{fig:phasespace}, we now consider the properties
of the final planetary system for the outcomes which yield a planetary 
system---namely resonant preservation, and exchange of the host star for the incoming star.
Since resonance is only allowed for $v_\infty < v_c$, for simplicity we consider
only the outcome ``ex$\_$s (non-res)'', which has a significant cross section 
for all $v_\infty$.  Shown in Figure~\ref{fig:hardsoft} are the statistical properties
of the final planetary systems in the outcome ``ex$\_$s (non-res)'' as a function of $v_\infty/v_c$, for the case
$M_1=M_2=M$, $q=10^{-3}$ and $e=0$.  The fraction of final planetary systems that have 
hardened (semimajor axis decreased), $f_{\rm hard}$, is shown in the solid line.  
The average change in the binding energy of the planetary system, 
$\langle \Delta E_b\rangle/E_b=\langle GMm/2a^\prime- GMm/2a\rangle/(GMm/2a)=a\langle 1/a^\prime\rangle-1$, 
is shown in the dashed line.  The vertical dotted lines are the critical velocity and the orbital 
speed.  The horizontal dotted line represents the value 0.5 for $f_{\rm hard}$ and 0 
for $\langle \Delta E_b \rangle/E_b$.  For $v_\infty \ll v_c$, every outcome results in the 
planetary system hardening, with $\langle \Delta E_b \rangle/E_b \approx 0.4$.  This implies that
the semimajor axis has shrunk to about 70\% of its initial value, as shown in Figure~\ref{fig:dascale},
which plots $\langle a/a^\prime\rangle^{-1}$ as a function of $v_\infty/v_c$.  In the
region $v_c < v_\infty < v_{\rm orb}$, the planetary system predominantly widens as a
result of the encounter, with only $\approx 30\%$ of systems hardening.  The systems don't widen
by much, on average decreasing their energy by about 20\%.  This corresponds to an increase
in the semimajor axis of about 25\%, as shown in Figure~\ref{fig:dascale}.  This is rather 
interesting, since in this region exchange is nearly as likely as ionization, implying that a
significant fraction of systems that undergo a strong dynamical encounter survive it.  And those
that survive suffer a comparatively small change in their semimajor axis, with a significant
fraction actually hardening.  
Finally, in the region $v_\infty > v_{\rm orb}$,
since the cross section for ionization is several orders of magnitude larger than that of exchange,
the statistics are too poor to say anything conclusive.  Although the curves in Figures~\ref{fig:hardsoft}
and \ref{fig:dascale} appear to begin to rebound, we expect that they actually go smoothly
to $f_{\rm hard}=0$, $\langle \Delta E_b \rangle / E_b=-1$, and 
$\langle a/a^\prime \rangle^{-1}\rightarrow\infty$.

As described in section~\ref{sec:analytic}, there has been considerable variance in the literature
on what velocity demarcates the hard--soft boundary for systems with a wide mass ratio.  
A look at Figure~\ref{fig:hardsoft} reveals that (at least for $M_1=M_2$), the hard--soft 
boundary---which lies where $f_{\rm hard}$ crosses $0.5$ and $\langle \Delta E_b \rangle/E_b$ crosses 
$0$---robustly occurs at $v_\infty \approx 0.75 v_c$.  Of course, simply stating that the hard--soft boundary
lies at $v_\infty \approx 0.75 v_c$ belies the fact that for $v_c < v_\infty < v_{\rm orb}$
the likelihood of a planetary system surviving a strong dynamical encounter in some
form is rather significant.

\begin{figure}
  \begin{center}
    \includegraphics[width=\columnwidth]{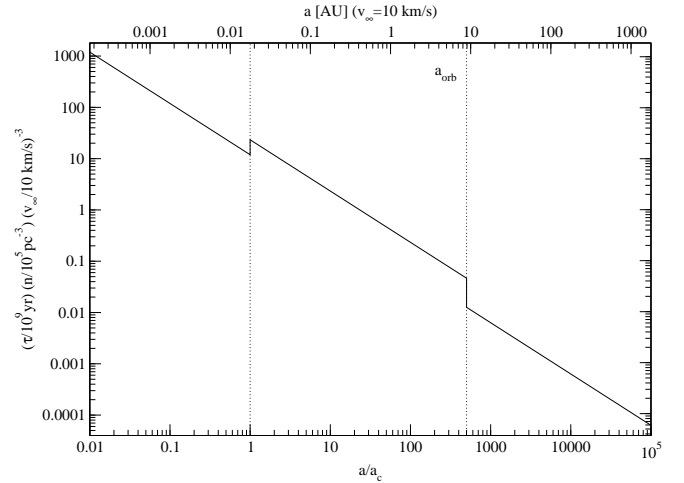}
    \caption{The characteristic lifetime $\tau$, plotted in the quantity $\tau n/v_\infty^3$
      (see eq.~[\ref{eq:taugeneral}]),
      of a planetary system
      with mass ratio $q=10^{-3}$, host star mass $M=M_\sun$, and eccentricity $e=0$ in a stellar environment of 
      number density $n$ and velocity dispersion $v_\infty$, as a function of $a/a_c$.
      The critical semimajor axis $a_c$ is as defined in section \ref{sec:hardsoft} after eq.~(\ref{eq:taugeneral}).  
      The two vertical
      dotted lines show $a=a_c$ and $a=a_{\rm orb}$.  The quantity $a_{\rm orb}$
      is the value of $a$ for which the orbital speed in the planetary system
      is equal to $v_\infty$.\label{fig:tau}}
  \end{center}
\end{figure}

Based on the understanding garnered from our analysis thus far, it is useful to consider 
the fate of a planetary system subject to strong dynamical encounters in a stellar 
environment of number density $n$ and velocity dispersion $v_\sigma$.  The average
relative velocity between objects in such a system is roughly $\sqrt{2}v_\sigma$, but since
we are interested only in orders of magnitude here we will take $v_\infty=v_\sigma$.
Since we know all relevant cross sections, we can perform a simple ``$n\sigma v$''
estimate for the lifetime of a planetary system as a function of $a$,
\begin{equation}\label{eq:tau}
  \tau = \frac{1}{n\sigma v_\infty} \, ,
\end{equation}
which we do now.
A useful way to think about the evolution of a planetary system subject to encounters is
the following.  Taking the masses to be given, for a given value of $a$, 
we know $v_c$ and $v_{\rm orb}$ from eqs.~(\ref{eq:vcrit})
and (\ref{eq:vorb}).  Given the value of $v_\infty$ fixed by the velocity dispersion of
the stellar environment in which the planetary system finds itself, we thus 
have $v_\infty/v_c$ and know in which section of parameter space the planetary system
lies in Figure~\ref{fig:sigmas}.  In the region $v_\infty < v_c$, by far the most 
likely outcome of a strong dynamical
interaction is exchange in which the planet is ejected.  Thus $\tau$ for this region
can be calculated using the cross section 
$\sigma_{\rm ex\_p}=\sigma_{\rm ex\_p (res)}+\sigma_{\rm ex\_p (non-res)}$.
In the region $v_\infty > v_{\rm orb}$ the dominant outcome is 
ionization, so one uses $\sigma_{\rm ion}$ in eq.~(\ref{eq:tau}).  
In between, in $v_c < v_\infty < v_{\rm orb}$, the outcomes
``ion'' and ``ex$\_$s (non-res)'' are almost equally likely.  From eqs.~(\ref{eq:i})
and (\ref{eq:j}), roughly $X_i \equiv 2/3$ of strong dynamical interactions result in
ionization, while the remaining $X_e \equiv 1/3$ result in non-resonant exchange of the
host star for the incoming star.  The systems that undergo exchange suffer only a small
perturbation to their orbital size, as evidenced by Figure~\ref{fig:dascale}.  As $a$
changes, $v_c$ and $v_{\rm orb}$ scale as $a^{-1/2}$ with $v_c/v_{\rm orb}$ fixed.
Thus, in effect, as $a$ changes, the quantity $v_\infty/v_c$ scales as $a^{1/2}$ with
the distance between $v_c$ and $v_{\rm orb}$ held fixed in log-space,
and so the planetary system simply occupies a new position on the $x$-axis in Figure~\ref{fig:sigmas}.
A planetary system lying in regions $v_\infty < v_c$ or $v_\infty > v_{\rm orb}$ is
destroyed as a result of an interaction.  But for a planetary system in the intermediate
region, it is destroyed a fraction $X_i$ of the time, while the other fraction $X_e$
of the time it survives with $v_\infty/v_c$ increased by a mere $\sim 10\%$ on average
(from Figure~\ref{fig:dascale}).  Eventually systems that survive via many exchange
interactions will find themselves in the region $v_\infty > v_{\rm orb}$ and be destroyed
quickly by ionization.  Since the average change in $a$ due to these exchanges
is rather small, let us, for the sake of simplifying the calculation of $\tau$, assume 
that they survive indefinitely.  The average lifetime in this region of parameter space,
including both ionization and exchange, is thus
\begin{equation}
  \tau = \tau_i + X_e \tau_i + X_e^2 \tau_i + X_e^3 \tau_i + \cdots = \frac{\tau_i}{1-X_e}\, ,
\end{equation}
where
\begin{equation}
  \tau_i = \frac{1}{n (\sigma_{\rm ion} + \sigma_{\rm ex\_s (non-res)}) v_\infty} \, .
\end{equation}
Since all cross sections of interest here scale as $\sigma = \sigma_0 a/v_\infty^2$ where
$\sigma_0$ is a function of the masses, we can be quite general and plot 
\begin{equation}\label{eq:taugeneral}
\frac{\tau n}{v_\infty^3} = \frac{1}{\sigma_0}\left(\frac{a}{a_c}\right)^{-1}\frac{\mu}{G M_1 m} \, ,
\end{equation}
where the critical semimajor axis is $a_c=GM_1m/\mu v_\infty^2$, 
as a function of $a/a_c$ in Figure~\ref{fig:tau}.
Note that the sharpness of the transitions at $a=a_c$ and $a=a_{\rm orb}$ is due to the use 
of the analytical cross sections in calculating the lifetimes.  From the numerical cross
sections in Figure~\ref{fig:sigmas} it is evident that, in reality, the transitions
are a bit smoother, although the general structure is the same.
Clearly the overall trend
is for the lifetime to scale as $a^{-1}$, with the coefficient varying by only 
a factor of a few among the different regions of parameter space.  
Using only the fact that the hard--soft boundary lies at $v_\infty=v_c$, one might
expect the lifetime to drop markedly at $a=a_c$.  In fact, it increases slightly
there, and does not drop until $a=a_{\rm orb}$, due to the large cross section
for planetary system preserving exchange in the intermediate regime.  We now consider
the implications of this result.

\section{Implications}\label{sec:implications}

The main implication of our results is that the minimum size of planetary 
systems that will likely be destroyed as a result of dynamical interactions
in dense stellar environments is larger than has previously been assumed.
This limit can be read almost directly from Figure~\ref{fig:tau}
for a given velocity dispersion, stellar density, and timescale.
For example, in a globular cluster like 47 Tuc, which has a central stellar
density of $\sim 10^5\,{\rm pc}^{-3}$ and velocity dispersion of 
$\sim 10\,{\rm km}/{\rm s}$, planetary systems wider than $0.05\,{\rm AU}$
will have been destroyed after $\sim 10\,{\rm Gyr}$.
The second implication is that the statement that the location of the hard--soft boundary
is at $v_\infty=v_c$, although correct, when taken on its own can be misleading.  It is true
that planetary systems on the low-velocity side of the hard--soft boundary
that survive interactions tend to harden on average.  However, planetary
systems are overwhelmingly likely to be destroyed in this regime, as can be
read directly from the cross sections in Figure~\ref{fig:sigmas}.  Similarly,
although planetary systems tend to soften, on average, for $v_\infty > v_c$,
it is not until $v_\infty > v_{\rm orb}$ that ionization dominates the outcome.
In other words, there is a wide region in parameter space (from $v_\infty = v_c$
to $v_\infty = v_{\rm orb}$) for which the probability of exchange in which a 
planetary system results is comparable that of ionization, and those resulting
planetary systems soften by only a small amount (as shown in 
Figures~\ref{fig:hardsoft} and \ref{fig:dascale}).

Our results have a bearing on previous studies in the literature in which 
authors have adopted inappropriate prescriptions for outcomes of scattering interactions
with planetary systems.  As an example, \citet{2001MNRAS.322..859B} calculated 
the evolution of a population of planetary
systems in stellar environments of differing density, using a Monte-Carlo
approach in which planetary systems undergo binary--single interactions with field
stars.  As a result of each interaction, the planetary system is assumed
to be ionized if $v_\infty > v_c$, and to harden if $v_\infty < v_c$.  
However, as our results show, for $v_\infty < v_c$ those planetary systems
undergoing strong interactions are destroyed, for $v_c < v_\infty < v_{\rm orb}$
they have an increased probability of surviving, and for $v_\infty > v_{\rm orb}$
they are predominantly ionized.  The result of the discrepancy
is that most of the systems destroyed on the high-velocity
side of the hard--soft boundary ($v_\infty = v_c$) in these simulations
will actually survive as
planetary systems of some kind.  The growth of a spike in the relative frequency
of planetary systems smaller than $a_c$ predicted in low velocity
dispersion systems (such as open clusters), should not occur, 
with the distribution of semimajor
axes dropping smoothly from $a_c$ to $a_{\rm orb}$.  There is the possibility
for a slight hump in the relative frequency of planets with $a_c < a < a_{\rm orb}$,
based on our Figure~\ref{fig:tau}, but it would not be nearly as pronounced as what
is shown in Figure 5 of \citet{2001MNRAS.322..859B}.

On the observational side, we find that of the hot Jupiter planetary systems 
expected to be detected in 47 Tuc in the transit survey of 
\citet{2000ApJ...545L..47G}---which was sensitive to periods up to $\approx 4\,{\rm d}$---essentially
all are expected to survive dynamical interactions in the high density
core of the cluster on timescales comparable to the age of the cluster.  This can be read
directly from Figure~\ref{fig:tau} with $a\approx 0.05\,{\rm AU}$.  
This bolsters the conclusion drawn by \citet{2005ApJ...620.1043W} that it
is not dynamics that is responsible for the lack of hot Jupiters detected in 47 Tuc,
but rather may be metallicity-dependent effects \citep{2005ApJ...622.1102F}.  
The \citeauthor{2005ApJ...620.1043W} study sampled regions outside the core
and was sensitive to orbital periods up to $\approx 16\,{\rm d}$, so the hot Jupiter 
binaries they looked for would also not 
have been destroyed by dynamical interactions due to the lower density there.  However,
it is possible that they would have been destroyed early in the lifetime of the cluster
if it went through a high-density phase early on, the possibility of which
is discussed in \citet{2001MNRAS.322..859B}.

Looking ahead, we now speculate on future possibilities for observations of planets
in dense stellar environments.  The globular cluster Terzan 5 is unique in that it
has super-solar metallicity.
Any argument against the formation of planets in globular clusters due to metallicity
effects is null in Terzan 5.  If we adopt the frequency of hot Jupiters in the solar
neighborhood of 0.8\% \citep{2000ApJ...545..504B,2004Natur.430...24B}, and use
the estimated central number density of $10^5\,{\rm pc}^{-3}$ and core radius
of $0.5\,{\rm pc}$ \citep{1996AJ....112.1487H}, we expect $\sim 100$ hot 
Jupiters in orbit around main sequence stars in the core of Terzan 5.
According to Figure~\ref{fig:tau} for $v_\infty=10\,{\rm km}/{\rm s}$, 
all of these systems would survive for a Hubble time, since
the definition of ``hot Jupiter'' implies an orbit period
less than $\sim 10\,{\rm d}$.  Terzan 5 lies nearly in the Galactic disk, 
close to the Galactic center, making optical observations of it practically impossible.  
However, it is readily observed in radio wavelengths, and already many millisecond
pulsars (MSPs) have been discovered in it \citep{2005Sci...307..892R}.  There is thus
the possibility of detecting planets in orbits around MSPs in this cluster.  The
most natural formation mechanism involves exchanging the MSP for the planet's
host star in a dynamical interaction.  From Figure~\ref{fig:phasespace} we
see that for the velocity dispersion of Terzan 5 planetary systems with 
$a \sim 0.01$--$10\,{\rm AU}$ lie in the intermediate regime, and are thus
most likely to possibly have exchanged in a MSP.  The formation of such systems
is a balancing act between destruction (for the wider systems) and a long interaction
timescale (for the smaller systems), with a ``sweet spot'' at $a \sim 0.1\,{\rm AU}$.
Although the expected number of MSP-planet systems is small (probably $\sim 1$), 
we can predict that such a system would have $a \sim 0.1\,{\rm AU}$.  There is 
also the possibility that planets could form around pulsars, but the frequency
of formation is rather less certain \citep{1992ApJ...399L..95S}.

\section{PSR B1620-26}\label{sec:1620}

The planetary system PSR B1620-26 in the globular cluster M4 consists
of a Jovian-mass object in an $\approx 20\,{\rm AU}$ orbit about
a MSP--WD binary.  The planetary system
sits right at the edge of the core of M4 in projection, which has a central
velocity dispersion of $\approx 4\,{\rm km}/{\rm s}$.  Looking at 
Figure~\ref{fig:phasespace}, we see that the planetary system
lies comfortably in the ``intermediate'' regime, making it very likely
that it was formed via a dynamical scattering interaction in which the 
NS-containing binary exchanged into the planetary system (as one unit) in place 
of the planet's original host star (scenario ``2'' below).  
With this motivation, we now perform a detailed numerical analysis of the 
possible formation scenarios of PSR B1620-26.

First, it is useful to list the physical parameters of the system.
The pulsar has a spin period of $\approx 11\,{\rm ms}$ and mass
$\approx 1.35\,M_\sun$, while the WD has mass $0.34 \pm 0.04\,M_\sun$.
Hubble Space Telescope data has confirmed that the WD is young, with an age of 
$\sim 0.5\,{\rm Gyr}$.  
The PSR--WD binary has eccentricity $\sim 0.025$.  The planet has mass 
$\approx 1$--$3 M_J$, where $M_J$ is the mass of Jupiter, 
and orbits with eccentricity $\sim 0.3$ about the MSP--WD binary.  
The inclination between the inner and outer orbits is $\sim 55^{\circ}{}^{+14}_{-8}$.  
The currently favored dynamical formation scenarios are the following.
In what we will call scenario ``1'', a binary composed of a WD and a NS
undergoes an exchange 
interaction with a binary consisting a main-sequence star near the turn-off mass 
and a Jovian planet.  In the encounter the WD is ejected from the 
system with the normal star exchanging into orbit with PSR, leaving the planet
in a wide orbit about the star--PSR binary.  The star then undergoes mass
transfer as it goes through the red giant phase, recycling the PSR to
millisecond periods \citep{2003Sci...301..193S}.
In scenario ``2'', a binary composed of a NS and a near turn-off main-sequence star
of mass $0.8\,M_\sun$ undergoes an exchange interaction with a binary composed of a main-sequence
star and a Jovian planet.  In the encounter the NS--star binary acts as a 
single unit and exchanges into the planet orbit, ejecting the planet's original host star.
The star then undergoes mass transfer as it goes through the red giant phase, 
recycling the PSR to millisecond periods \citep{2000ApJ...528..336F}.
As we discuss below, our expectation that scenario ``2'' is the much
more likely one proves to be correct.

\subsection{Quantifying the Outcome Likelihoods}

\begin{figure}
  \begin{center}
    \includegraphics[width=\columnwidth]{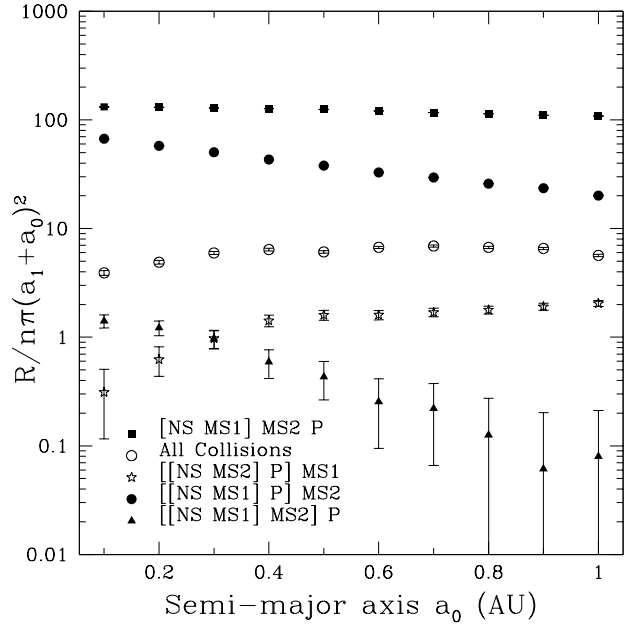}
    \caption{Quantifying the likelihood of the possible formation scenarios of PSR B1620-26: 
      The semi-normalized rate $R/[n\pi(a_0+a_1)^2]$ (in units of ${\rm km}/{\rm s}$)
      for various outcomes of a scattering interaction between a NS--MS1 binary
      and a MS2--P planetary system as a function of $a_0$, the semi-major axis
      of the NS--MS1 binary.  Points are as defined in the legend.  The rate 
      for scenario ``2'' (``[[NS MS1] P] MS2'') clearly dominates over 
      scenario ``1'' (``[[NS MS2] P] MS1'') for a wide range in $a_0$.\label{fig:sourav1}}
  \end{center}
\end{figure}

As in section~\ref{sec:numeric}, we use the \Fewbody\ numerical toolkit
to perform scattering interactions and calculate cross sections for different
outcomes.  The interaction rate per planetary system is simply
\begin{equation}
  R = \int_0^\infty dv_\infty f(v_\infty) n \sigma v_\infty \, ,
\end{equation}
where $f(v_\infty)$ is the velocity distribution function, $n$ is the number density
of NS binaries, and $\sigma$ is the cross section for the outcome of interest.  
Rates presented below were calculated by discretizing the integral
over the distribution function and summing over a range of velocities
wide enough to guarantee convergence.  In particular, 
we typically integrate over a range of velocities from $1\,{\rm km}/{\rm s}$
to $15\,{\rm km}/{\rm s}$, with the upper limit set by the escape speed of
$15.25\,{\rm km}/{\rm s}$ of the cluster \citep{2004MNRAS.355.1244B}.
For the distribution
function we assume a simple Maxwellian with a velocity dispersion of
$v_\infty=\sqrt{2} v_\sigma$, where $v_\sigma=3.88\,{\rm km}/{\rm s}$ 
is the velocity dispersion in the core of M4 and the factor of $\sqrt{2}$ is due to the
fact that we are considering relative velocities between pairs of objects.
Our naming convention for the objects is analogous to what we used above, 
with ``[NS MS1] [MS2 P]'' denoting the
initial configuration of the two binaries before the scattering
interaction.  ``MS1'' in fact represents a WD in scenario ``1'' and 
a main-sequence star in scenario ``2''; however, we label it ``MS1'' in
both cases since it is only the mass of the object that matters in the scattering
interactions (the radius
of the object plays only a minor role when direct physical collisions are considered).
We label the semimajor axis of the NS--MS1 binary $a_0$, and that of the
MS2--P binary $a_1$.  Since the NS binary must be significantly tighter than 
the planet binary, it is expected to behave dynamically as a single unit; thus
the cross sections for different outcomes are expected to scale
with $a_1$ (our numerical results confirm this).  We therefore vary only $a_0$,
from $0.1\,{\rm AU}$ to $1\,{\rm AU}$ (a typical range of values), and keep
$a_1$ fixed at $5\,{\rm AU}$.  We fix $e_1$, the eccentricity of the planet binary,
at 0, while we draw $e_0$, the eccentricity of the NS binary from a thermal
distribution truncated at large $e_0$ such that neither object in the binary
exceeds its Roche lobe.  
Figure~\ref{fig:sourav1} shows the semi-normalized rate $R/[n\pi(a_0+a_1)^2]$
(in units of ${\rm km}/{\rm s}$) for the different outcomes listed in the legend as a function of $a_0$.
The most likely outcome is ``[NS MS1] MS2 P'', i.e., destruction of the planetary
system with preservation of the NS binary.  The next most likely outcome
is scenario ``2'' (``[[NS MS1] P] MS2''), confirming our expectations from above.
The next is any outcome involving direct physical collisions between system
members, which we lump together into one category for the sake of succinctness.
The next is scenario ``1'' (``[[NS MS2] P] MS1'') which is larger than
``[[NS MS1] MS2] P'' only for $a_0 \ga 0.3\,{\rm AU}$.

\begin{figure}
  \begin{center}
    \includegraphics[width=\columnwidth]{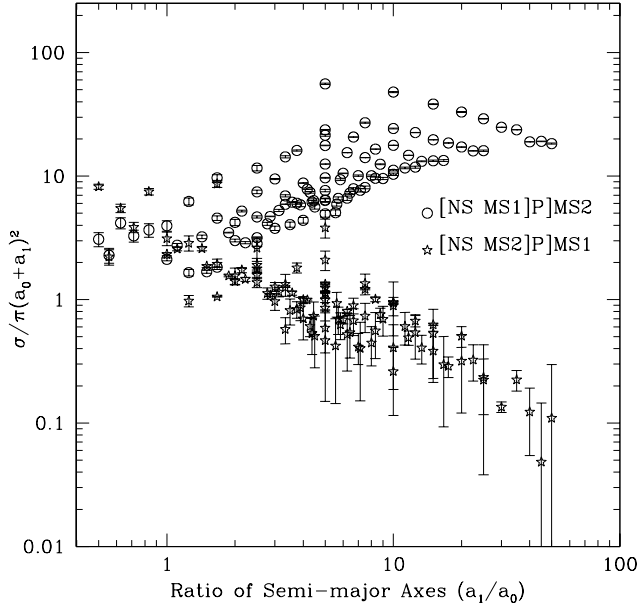}
    \caption{The normalized cross section, $\sigma/\pi(a_0+a_1)^2$, for PSR B1620-26 formation scenarios
      ``1'' (stars) and ``2'' (circles) as a function of $a_1/a_0$, where $a_0$ is the initial
      semimajor axis of the neutron star binary and $a_1$ is that of the planet binary.  The cross sections
      are calculated over a large grid in $a_0$--$a_1$ space, hence their multi-valued
      nature as a function of $a_1/a_0$.  All interactions were performed with 
      $v_\infty=5\,{\rm km}/{\rm s}$ since the velocity-dependent rate peaks 
      at this value.  It is clear from the figure that as long as $a_1 \ga a_0$,
      which is quite reasonable on physical grounds, scenario ``2'' is the most 
      likely.\label{fig:sourav2}}
  \end{center}
\end{figure}  

To verify that our results are not dependent on the choice of $a_1$,
we now consider the relative probabilities for scenarios ``1'' and ``2''
over a large grid in $a_0$--$a_1$ space.  To simplify the presentation
of the results we plot the normalized cross section for each outcome,
$\sigma/\pi(a_0+a_1)^2$, as a function of $a_1/a_0$, as shown in
Figure~\ref{fig:sourav2}, hence the multi-valued nature of the cross
sections.  For simplicity all interactions were performed with 
$v_\infty=5\,{\rm km}/{\rm s}$ since the velocity-dependent rate peaks 
at this value.  It is clear from the figure that as long as $a_1 \ga a_0$,
which is quite reasonable on physical grounds, scenario ``2'' is the most 
probable.  For the physically most reasonable value of $a_1/a_0 \approx 10$, 
scenario ``2'' is more probable by more than an order of magnitude over 
scenario ``1''.

\begin{figure}
  \begin{center}
    \includegraphics[width=\columnwidth]{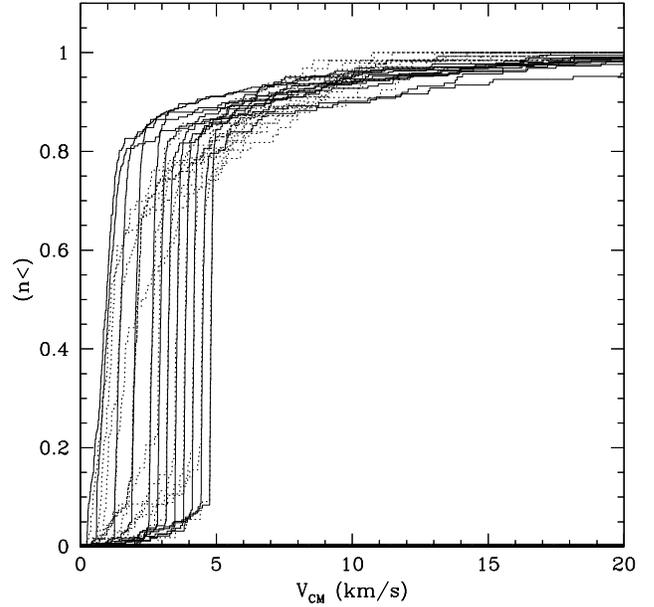}
    \caption{The cumulative distribution of the center-of-mass
      velocity of the putative PSR B1620-26 triple after the exchange interaction in scenario ``2'', 
      for the cases $a_0=0.1\,{\rm AU}$
      (solid lines) and $a_0=1\,{\rm AU}$ (dotted lines) for several values of 
      $v_\infty$ from $1\,{\rm km}/{\rm s}$ (left-most) to $15\,{\rm km}/{\rm s}$ (right-most).  
      In both cases we set $a_1=5\,{\rm AU}$.
      The triple system appears to suffer a recoil of $\sim 3\,{\rm km}/{\rm s}$, which is clearly
      not enough to eject it from the cluster (requiring $\approx 15\,{\rm km}/{\rm s}$), 
      but may be enough to displace it significantly away from the core.\label{fig:sourav3}}
  \end{center}
\end{figure} 

\begin{figure}
  \begin{center}
    \includegraphics[width=\columnwidth]{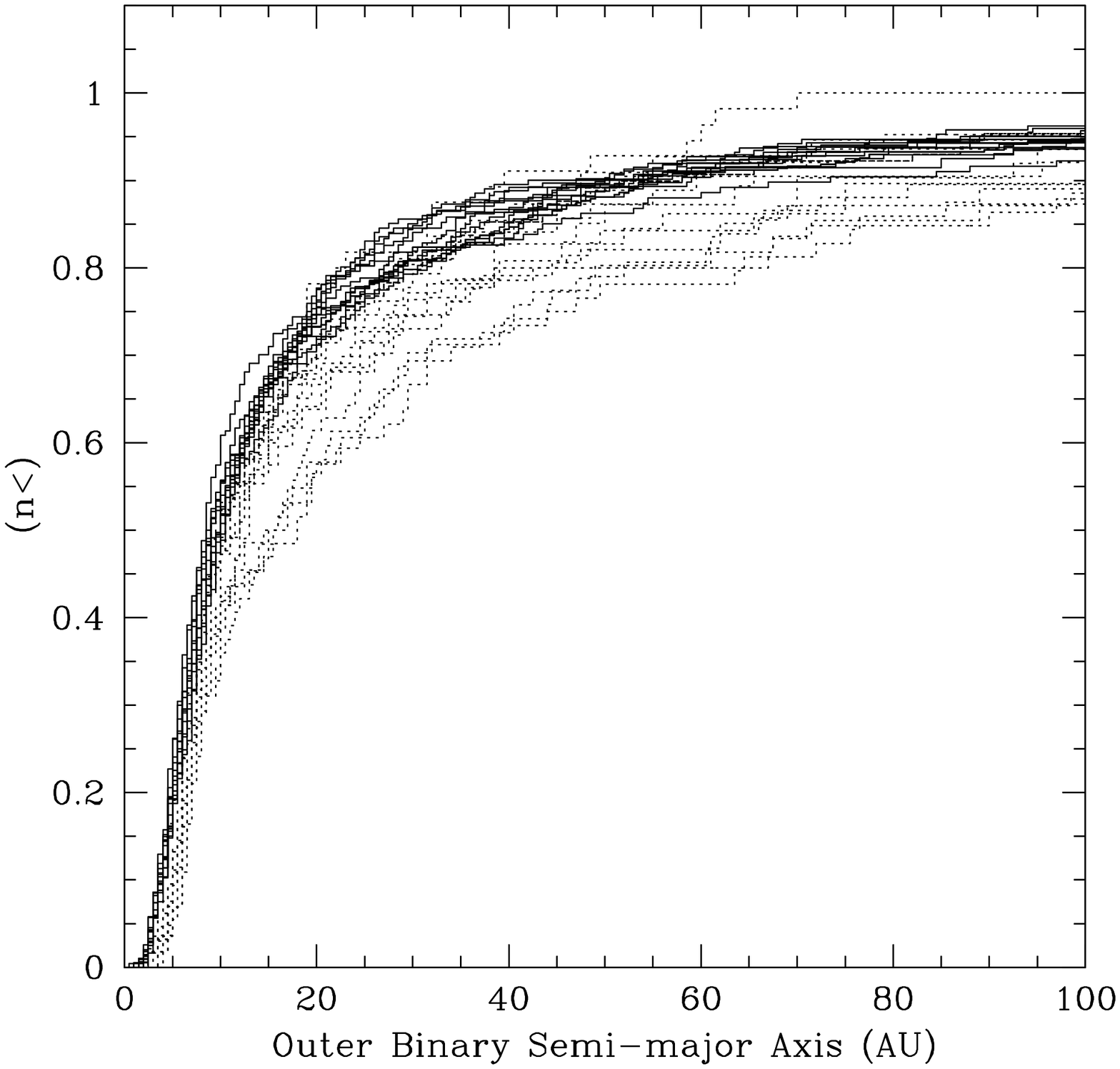}
    \caption{The cumulative distribution of the outer semimajor axis, $a_{\rm out}$,
      for the putative PSR B1620-26 triple system formed in scenario ``2''.  Line conventions and initial conditions 
      are the same as in Figure~\ref{fig:sourav3}.\label{fig:sourav4}}
  \end{center}
\end{figure}  

\begin{figure}
  \begin{center}
    \includegraphics[width=\columnwidth]{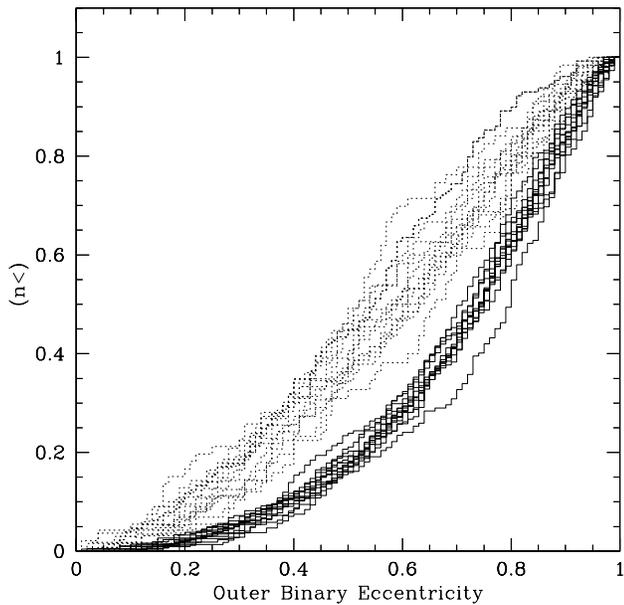}
    \caption{The cumulative distribution of the outer eccentricity, $e_{\rm out}$,
      for the putative PSR B1620-26 triple system formed in scenario ``2''.  Line conventions and initial conditions 
      are the same as in Figure~\ref{fig:sourav3}.\label{fig:sourav5}}
  \end{center}
\end{figure}  

\begin{figure}
  \begin{center}
    \includegraphics[width=\columnwidth]{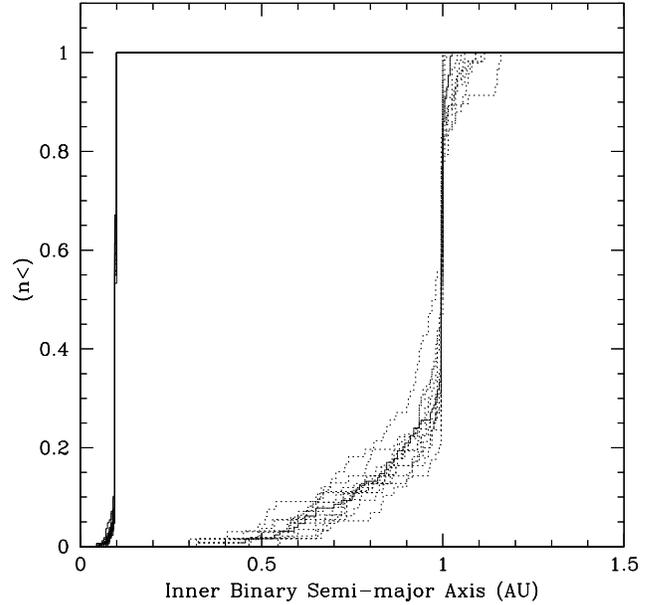}
    \caption{The cumulative distribution of the inner semi-major axis, $a_{\rm in}$,
      for the putative PSR B1620-26 triple system formed in scenario ``2''.  Line conventions and initial conditions 
      are the same as in Figure~\ref{fig:sourav3}.\label{fig:sourav6}}
  \end{center}
\end{figure}  

Having confirmed the preferred formation scenario, we now consider
the orbital parameters of the hierarchical triple formed.
Figure~\ref{fig:sourav3} shows the cumulative distribution of the center-of-mass
velocity of the triple after the exchange interaction, for the cases $a_0=0.1\,{\rm AU}$
(solid lines) and $a_0=1\,{\rm AU}$ (dotted lines) for several values of 
$v_\infty$ from $1\,{\rm km}/{\rm s}$ (left-most) to $15\,{\rm km}/{\rm s}$ (right-most).  The triple
system appears to suffer a recoil of $\sim 3\,{\rm km}/{\rm s}$, which is clearly
not enough to eject it from the cluster (requiring $\approx 15\,{\rm km}/{\rm s}$), 
but may be enough to eject it from the core.  
Figures~\ref{fig:sourav4} and \ref{fig:sourav5} show the cumulative
distributions of outer semimajor axis, $a_{\rm out}$, and eccentricity,
$e_{\rm out}$, respectively, with the same convention
for the line types as in the previous figure.  The outer semimajor axis is clearly
not particularly dependent on either $a_0$ or $v_\infty$, 
with a peak at $\sim 10\,{\rm AU}$.  The outer eccentricity, on the other hand,
while not particularly dependent on $v_\infty$, is strongly dependent on $a_0$.
For $a_0=0.1\,{\rm AU}$ the eccentricity distribution is roughly thermal, with
the cumulative distribution $\propto e_{\rm out}^2$.  For $a_0=1\,{\rm AU}$ the eccentricity
distribution is flatter, with more systems at lower
eccentricity.  This result suggests that of the two, $a_0=1\,{\rm AU}$ is the favored
value for the NS binary semimajor axis, since we know from observations
that the outer eccentricity is rather low, with 
$e_{\rm out} \sim 0.2 \pm 0.1$ \citep{psr:update}.
In Figure~\ref{fig:sourav6} we show the cumulative distribution of the
inner binary semimajor axis, $a_{\rm in}$, with the same convention for 
the line types as in the previous figures.  Confirming our earlier assumption,
the inner binary clearly behaves as a single dynamical unit, with the 
distributions for the final values strongly peaked around their initial 
values, and with very minimal dependence on $v_\infty$.

\subsection {Evolution of the Hierarchical Triple Due to Mass Transfer}

In scenario ``2'', the hierarchical triple formed via the dynamical interaction is not the
planetary system we see today in M4, but merely its progenitor.  In this section
we calculate the effects on the triple due to mass transfer in the inner binary 
as the main-sequence star ascends the red giant branch, and apply the results
to the ensemble of triples formed in our scattering interactions.  
We assume slow isotropic mass loss, and that during
this process the eccentricities remain invariant \citep{1992A&A...256L..35R}.
For the sake of simplifying the calculation we also assume that the inner
and outer eccentricities are zero.
Since the outer orbit carries so little angular momentum ($\sim 0.4\%$ of the total angular
momentum of the triple system) we neglect angular momentum coupling between the 
two orbits and treat them as two separate binaries during the mass transfer.

The angular momentum of the inner orbit is given by
\begin{equation}\label{eq:sourav6}
L_{\rm in} = M_{\rm NS} M_{\rm MS1} \left(\frac{Ga_{\rm in}}{M_{\rm in}} \right)^{1/2} \, ,
\end{equation}
where $M_{\rm in}=M_{\rm NS}+M_{\rm MS1}$, $M_{\rm NS}$ and $M_{\rm MS1}$ are the masses of the NS
and the WD progenitor star, respectively, and $a_{\rm in}$ is the semimajor
axis of the inner binary.  Similarly, the angular momentum of the outer orbit is given by
\begin{equation}\label{eq:sourav7}
L_{\rm out } = M_{\rm in} M_{\rm P} \left(\frac{Ga_{\rm out}}{M_T}\right)^{1/2} \, , 
\end{equation} 
where $M_T=M_{\rm in}+M_{\rm P}$, $M_P$ is the mass of the planet, and 
$a_{\rm out}$ is the semimajor axis of the outer orbit.
Taking the logarithm of both sides and differentiating, we get
\begin{equation}\label{eq:sourav8}
\frac {\delta L_{\rm in}}{L_{\rm in}}=
\frac {\delta M_{\rm NS}}{M_{\rm NS}}+
\frac {\delta M_{\rm MS1}}{M_{\rm MS1}}+
\frac{1}{2}\frac{\delta a_{\rm in}}{a_{\rm in}}- 
\frac{1}{2}\frac{\delta M_{\rm in}}{M_{\rm in}}
\end{equation}
and
\begin{equation}\label{eq:sourav9}
\frac {\delta L_{\rm out}}{L_{\rm out}}=
\frac {\delta M_{\rm in}}{M_{\rm in}}+
\frac {\delta M_{\rm P}}{M_{\rm P}}+
\frac{1}{2}\frac{\delta a_{\rm out}}{a_{\rm out}}- 
\frac{1}{2}\frac{\delta M_T}{M_T}
\end{equation}
for differential changes in the inner and outer orbits.

In the mass transfer process, we assume that the mass lost
from MS1 gets channeled into an accretion disk around the NS,
and any mass lost from the system is lost from the accretion
disk with the characteristic angular momentum of the NS 
\citep[see e.g.][]{2002ApJ...565.1107P}.  This implies that
\begin{equation}\label{eq:sourav10}
\delta L_{\rm in}=a^2_{\rm in}\frac{M^2_{\rm MS1}}{M^2_{\rm in}}
\Omega_{\rm in}\delta M_{\rm in} \, ,
\end{equation}
where $\Omega_{\rm in}$ is the orbital velocity of NS, given by
\begin{equation} \label{eq:sourav11}
\Omega_{\rm in}=\left(\frac{GM_{\rm in}}{a^3_{\rm in}}\right)^{1/2} \, .
\end{equation}
Similarly, as mass is lost from the inner binary in the mass transfer process
the angular momentum of the outer binary will change
according to 
\begin{equation}\label{eq:sourav12}
\delta L_{\rm out}=a^2_{\rm out}\frac{M^2_{\rm P}}{M^2_T} \Omega_{\rm out}\delta M_T \, ,
\end{equation}
where $\Omega_{\rm out}$ is the orbital angular velocity of the inner binary
in the outer orbit, and is given by
\begin{equation}\label{eq:sourav13}
\Omega_{\rm out}=\left(\frac{GM_T}{a^3_{\rm out}}\right)^{1/2} \, .
\end{equation}

We parameterize the mass loss by defining $\beta$ as the fraction of
mass lost from MS1 that is accreted by the NS, so that
$\delta M_{\rm NS} = -\beta \delta M_{\rm MS1}$ and 
$\delta M_{\rm in} = (1-\beta) \delta M_{\rm MS1}$.  Note that with our conventions 
$\delta M_{\rm MS1} < 0$.
Combining equations (\ref{eq:sourav6}) through (\ref{eq:sourav13}), converting $\delta$'s to differentials
and integrating, we find \citep{2002ApJ...565.1107P}:
\begin{multline}\label{eq:sourav14}
  \frac{a_{\rm in}^\prime}{a_{\rm in}} = \left(\frac{M_{\rm MS1}^\prime}{M_{\rm MS1}}\right)^{-2} 
  \left(\frac{M_{\rm NS}^\prime}{M_{\rm NS}}\right)^{-2/\beta}
  \left(\frac{M_{\rm in}^\prime}{M_{\rm in}}\right)^{-1}\\
  \quad \mbox{for $0 < \beta \leq 1$} \, ,
\end{multline}
\begin{multline}\label{eq:sourav15}
  \frac{a_{\rm in}^\prime}{a_{\rm in}} = \left(\frac{M_{\rm MS1}^\prime}{M_{\rm MS1}}\right)^{-2} 
  \left(\frac{M_{\rm in}^\prime}{M_{\rm in}}\right)^{-1}\\
  \times\exp\left[2\left(\frac{M_{\rm MS1}^\prime-M_{\rm MS1}}{M_{\rm NS}}\right)\right]
  \quad \mbox{for $\beta=0$} \, ,
\end{multline}
and
\begin{equation}\label{eq:sourav16}
  \frac{a_{\rm out}^\prime}{a_{\rm out}} = \left(\frac{M_T^\prime}{M_T}\right)^{-1} 
  \quad \mbox{for all $\beta$} \, .
\end{equation}

\begin{figure}
  \begin{center}
    \includegraphics[width=\columnwidth]{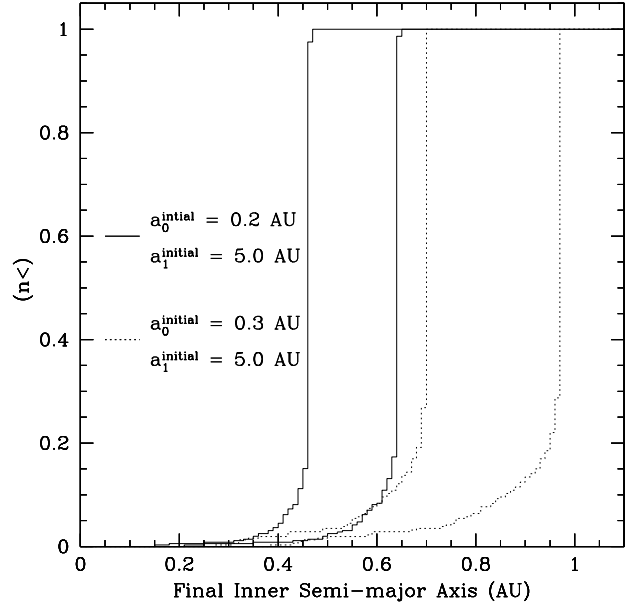}
    \caption{The cumulative distribution
      of the inner binary semi-major axis of the putative PSR B1620-26 triple system 
      after mass transfer.  The solid lines
      are for the case $a_0=0.2\,{\rm AU}$---in other words, the NS--MS1 semimajor axis was
      $0.2\,{\rm AU}$ before the scattering interaction.  The dotted lines
      are for the case $a_0=0.3\,{\rm AU}$.  In both cases we set $a_1=5\,{\rm AU}$ and
      $v_\infty=5\,{\rm km}/{\rm s}$.  For each set of the lines the rightmost represents
      completely non-conservative mass transfer ($\beta=0$), while the leftmost represents 
      completely conservative mass transfer ($\beta=1$).\label{fig:sourav7}} 
  \end{center}
\end{figure} 

\begin{figure}
  \begin{center}
    \includegraphics[width=\columnwidth]{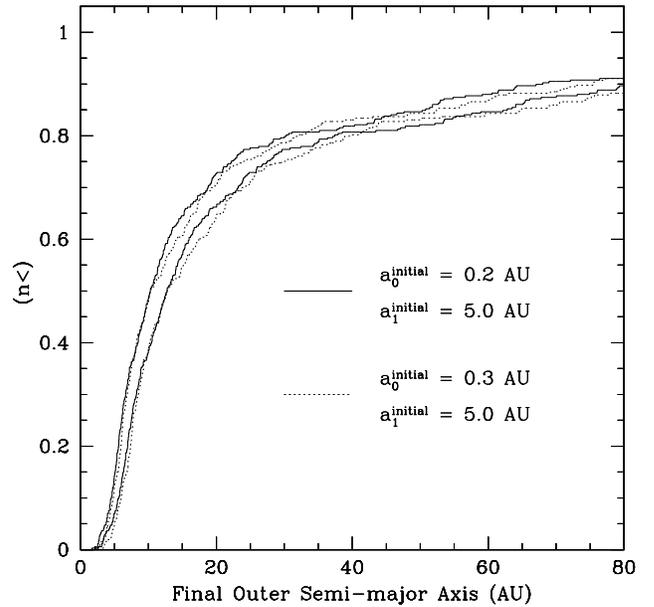}
    \caption{The cumulative distribution of the outer semi-major axis 
      of the putative PSR B1620-26 triple system after 
      mass transfer for the cases $a_0=0.2\,{\rm AU}$ (solid lines)
      and $a_0=0.3\,{\rm AU}$ (dotted lines).  Line conventions 
      and initial conditions are the same as in Figure~\ref{fig:sourav7}.\label{fig:sourav8}} 
  \end{center}
\end{figure}  

We can apply the expressions in eqs.~(\ref{eq:sourav14}) through (\ref{eq:sourav16})
to the ensemble of triples formed via scenario ``2'' to determine their properties
after mass transfer.  Figure~\ref{fig:sourav7} shows the cumulative distribution
of the inner binary semi-major axis after mass transfer.  The solid lines
are for $a_0=0.2\,{\rm AU}$---in other words, the NS--MS1 semimajor axis was
$0.2\,{\rm AU}$ before the scattering interaction.  The dotted lines
are for the case $a_0=0.3\,{\rm AU}$.  In both cases we set $a_1=5\,{\rm AU}$ and
$v_\infty=5\,{\rm km}/{\rm s}$.  For each set of the lines the rightmost represents
completely non-conservative mass transfer ($\beta=0$), while the leftmost represents 
completely conservative mass transfer ($\beta=1$).
Figure~\ref{fig:sourav8} shows the cumulative distribution of
the outer semi-major axis for the cases $a_0=0.2\,{\rm AU}$ and $a_0=0.3\,{\rm AU}$.  Line types are as
in the previous figure.  Comparing the distributions of the inner semimajor
axis after the scattering interaction in Figure~\ref{fig:sourav6} with those 
after mass transfer in Figure~\ref{fig:sourav7}, we see that the effect of mass transfer
is to create a spread in the distribution of order the initial value.  
From observations we know the period of the inner orbit to be $191\,{\rm d}$, corresponding
to a semimajor axis of $\approx 0.8\,{\rm AU}$ \citep{psr:update}.
From this figure we see that we can quite reasonably create such a system
with $a_0\approx 0.3\,{\rm AU}$ for a wide range in the mass transfer parameter $\beta$.
From Figure~\ref{fig:sourav8} we see that the distribution of the outer semimajor axis
after mass transfer is much broader.  Although the peak in the distribution appears
to occur around $5$--$10\,{\rm AU}$, it is quite plausible to form a system with an
outer semimajor axis of $20\,{\rm AU}$, as we observe for the system today.

\section{Summary and Conclusions}\label{sec:summary}

We have studied in detail dynamical scattering interactions
between star--planet systems and other single stars, tabulating
analytical cross sections for all possible outcomes of the
scattering encounter, and confirming them with numerical experiments (as shown in
Figure~\ref{fig:sigmas}).
In the scattering of planetary systems (or wide mass ratio binaries
in general) there are two characteristic velocities in the problem (as shown in 
Figure~\ref{fig:phasespace}):
the critical velocity $v_c$, which is the value of $v_\infty$ for which
the total energy of the three-body system is zero; and $v_{\rm orb}$,
the orbital speed in the binary.  For a mass ratio $q \la 10^{-3}$ the two characteristic
velocities differ by more than an order of magnitude, opening up a
region in parameter space in which the nature of the scattering process
is fundamentally different from what occurs in comparable mass
encounters.  For $v_c < v_\infty < v_{\rm orb}$ two outcomes dominate
the cross with roughly equal weight: ionization, which destroys the planetary
system; and exchange of the incoming star for the planet's original host star.
Thus in this region---which does not exist for comparable mass systems---there 
is a significant probability that a planetary system undergoing a strong dynamical
encounter survives, albeit with a new host star.  Since in the past workers in the field
have typically taken $v_c$ as the velocity above which planetary systems are 
predominantly destroyed, this implies greater survivability of planetary
systems in dense stellar environments than has been predicted in some cases,
as the characteristic lifetime plotted in Figure~\ref{fig:tau} shows.

We have also numerically determined the location of the hard--soft boundary for
planetary systems, and find that it lies at $v_\infty \approx 0.75 v_c$.  In other
words, interactions with $v_\infty < v_c$ in which the binary
survives (e.g., via exchange) typically harden ($a$ decreases), while
for $v_\infty > v_c$ surviving binaries soften ($a$ increases).  However,
this simple statement belies the rich structure of the parameter space.
In particular, for $v_c < v_\infty < v_{\rm orb}$, although the surviving
binary typically softens, it softens by only a small amount, 
increasing its semimajor axis by only $\sim 25\%$, as shown in Figure~\ref{fig:dascale}.  
Its final eccentricity, of course, is uncorrelated with the initial value, taking on a 
thermal distribution ($dP/de = 2e$).  Only for $v_\infty > v_{\rm orb}$ does the
binary soften by a large amount---that is, if it is not destroyed completely, which
is the overwhelmingly dominant outcome for this region of parameter space.

We have briefly explored the implications of our results.  As mentioned above,
the main implication is that the minimum size of a planetary system that will
likely be destroyed as a result of dynamical interactions in dense stellar 
environments is larger than has been previously been assumed.
For a stellar environment of a given velocity dispersion and density this 
limit can be read almost directly off of Figure~\ref{fig:tau}.  Applying our results
to the observational campaigns searching for hot Jupiters in 47 Tuc via stellar transits,
we find that the planetary systems that the \citet{2000ApJ...545L..47G} observations
were sensitive to would have in fact survived to the present day.  
This bolsters the conclusion drawn by \citet{2005ApJ...620.1043W} that it
is not dynamics that is responsible for the lack of hot Jupiters detected in 47 Tuc,
but rather may be metallicity-dependent effects.
Looking toward the future, we predict a MSP--planet binary may likely be detected
via pulsar timing in the super-solar metallicity cluster Terzan 5, with a semi-major
axis of $\sim 0.1\,{\rm AU}$.  

Recognizing that the planetary system PSR B1620-26 in the globular cluster M4
lies in the ``intermediate'' region of parameter space, we have performed
a detailed numerical study of the possible scenarios in which it may have
formed.  By calculating cross sections and interaction rates, and treating the effects
of mass transfer in the inner binary, we conclude
that the scenario in which a NS--MS binary exchanges
as a single dynamical unit into the planetary system for the planet's original host star 
\citep[similar to the scenario of][]{2000ApJ...528..336F}, is strongly favored 
over that of \citet{2003Sci...301..193S}, in which the exchange interaction involves
ejecting the original NS binary companion, with the planet's host star exchanging 
into it.  This statement is quantified in Figure~\ref{fig:sourav2} for a wide range
of initial binary semimajor axes.

Our general analysis (sections \ref{sec:analytic} through \ref{sec:implications}) 
does not consider the effects of direct physical collisions between
objects (e.g., MS stars and planets), which may play an important role when the ratio
of stellar size to binary semimajor axis is $\ga 0.01$ \citep{2004MNRAS.352....1F}.
However, our code accurately treats physical collisions, and we have allowed for collisions
in our analysis of PSR B1620-26 in section \ref{sec:1620}, as shown in the cross sections of
Figure~\ref{fig:sourav1}.  
We have not looked at the ejection speeds of the planets in encounters which destroy the
planetary system.  This topic has already been studied in detail using $N$-body
methods \citep{2001MNRAS.322L...1S,2002ApJ...565.1251H}.  

Since our results assume point-mass particles, they are applicable to systems
of any mass, as long as the mass ratio in the binary is relatively small $\la 0.1$,
and the incoming object and the heaviest object in the binary are of comparable
mass.  We have not yet explored the application to scenarios involving, e.g.,
intermediate-mass black holes and super-massive black holes, stars
and intermediate-mass black holes, and planetesimals and planets.  Our results
may also have limited application to wide mass ratio binaries containing more than
one light body, as long as the total mass in light objects is significantly less
than the host mass, and the orbits of the light objects are not so strongly
coupled that they transmit perturbations on timescales comparable to the orbital
period.

\acknowledgements

It is a pleasure to thank Natasha Ivanova for informative discussions about mass 
transfer in binaries.  We also thank the anonymous referee for many helpful comments on the manuscript.
This work was supported by NSF grants AST-0206182 and AST-0507727, and NASA
grant NAG5-12044.

\bibliographystyle{apj}
\bibliography{apj-jour,main,sourav}

\end{document}